# Characteristic Analysis of 1024-Point Quantized Radix-2 FFT/IFFT Processor

Rozita Teymourzadeh, *Member IEEE/IET*, Memtode Jim Abigo, Mok Vee Hong, *Member IET*
Faculty of Engineering, Technology & Built Environment,
UCSI University, 56000, Malaysia
rozita@ucsi.edu.my

***Abstract*-The precise analysis and accurate measurement of harmonic provides a reliable scientific industrial application. However, the high performance DSP processor is the important method of electrical harmonic analysis. Hence, in this research work, the effort was taken to design a novel high-resolution single 1024-point fast Fourier transform (FFT) and inverse fast Fourier transform (IFFT) processors for improvement of the harmonic measurement techniques. Meanwhile the project is started with design and simulation to demonstrate the benefit that is achieved by the proposed 1024-point FFT/IFFT processor. Pipelined structure is incorporated in order to enhance the system efficiency. As such, a pipelined architecture was proposed to statically scale the resolution of the processor to suite adequate trade-off constraints. The proposed FFT makes use of programmable fixed-point/floating-point to realize higher precision FFT.**

## I. Introduction

Discrete Fourier Transform (DFT) is amongst the most fundamental operations in digital signal processing. However, the widespread uses of DFTs make its computational requirements an important issue. The direct computation of the DFT requires approximately $N^2$ operations where N is the transform size. The breakthrough of Cooley-Tukey (CT) FFT comes from the fact that it reduces the complexity to an order of $N\log_2 N$ operations. The FFT is therefore an efficient algorithm to compute the DFT and its inverse (IDFT). It has several applications in the field of signal processing including the real-time processing of wireless time-domain and frequency-domain signals especially for use in Orthogonal Frequency Division Multiplexing (OFDM) systems such as Digital Video Broadcasting (DVB), Digital Subscriber Line (xDSL) and WiMAX (IEEE 802.16) [1-4]. These applications require large-point FFT processing, such as 1024/2048/8192-point, FFTs for multiple carrier modulation.

Many FFT algorithms based on the CT decomposition such as radix-$2^2$, radix-$2^3$, radix-4, radix-(4+2), prime-factor as well as split-radix algorithms, have been proposed using the complex mathematical relationship to reduce the hardware complexity [2-3]. For example, in [4] one butterfly unit is used for all computations and $N+N.\log_2 N$ clock cycles are required for the computation of the FFT. A second implementation approach is for speed demanding applications, where one butterfly unit is used for each decimation stage of a radix-2 FFT [5]. A pipeline architecture based on the constant geometry radix-2 FFT algorithm, which uses $\log_2 N$ complex-number multipliers (more precisely butterfly units) and is capable of computing a full N-point FFT in N/2 clock cycles has been proposed in 2009 [8]. All these developments have introduced their own disadvantages, in addition to the age-long finite word-length effects of digital circuitry [7-9]. This paper thus, uses the pipeline architecture [6] to propose a model for the analysis of important design constraints like the finite word-length effects and amount of resolution needed to achieve the appropriate SNR [8-10] for the desired design needs using the statistical tools for the analysis of a range of feasible resolution.

## II. Architecture Development

### *A. Algorithm development of the decimation-in-time (DIT) radix-p FFT*

The DFT of an N-point sequence x[n] is given by:

$$X[k] = \sum_{n=0}^{N-1} x[n] W_N^{kn} \quad \text{For k = 0, 1, 2,...,N-1} \tag{1}$$

Where $W_N = e^{-j\left(\frac{2\pi}{N}\right)}$.

Consider the general formula of the DIT Radix-p FFT as follows:

$$X\left[k + r\left(\frac{N}{p}\right)\right] = \sum_{n=0}^{\frac{N}{p-1}}\left(\sum_{j=0}^{p-1} x[pn+j].W_p^{jr} W_N^{jk}\right) W_{\frac{N}{p}}^{nk} \tag{2}$$

for k = 0,1,2,…,N/p-1 and R = 0,1,2,…,p-1. Using the above decomposition, the DFT can be reduced successively to *N/p* p-point DFTs. In general, this process can be repeated m times and therefore there are totally *m* stages in the implementation of the DFT.

### *B. Parallel Architecture*

The computational structure of a butterfly unit is shown in Fig. 1. It is the fundamental computational of the parallel architecture

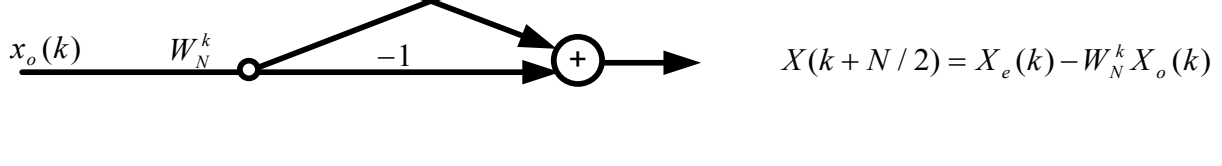

Fig. 1. Radix-2 butterfly unit

The butterfly unit requires a complex multiply and two complex additions. Therefore, it takes a total of $(N/2)\ \mathrm{Log}_2 N$ complex multiplies and $N\log_2 N$ complex additions to compute all $N$-point DFT samples. An 8-point Radix-2 DIT FFT requires $N/2$ butterfly units per stage for all $m$ stages [11-15]. For larger butterflies ($N > 2^6$), the processor becomes extremely complex and slow. Hence, a simpler and faster architecture is then required. Therefore, the proposed system was designed and simulated by MATLAB software. Fig. 2, shows the overall pipelined system structure and its designed control signals.

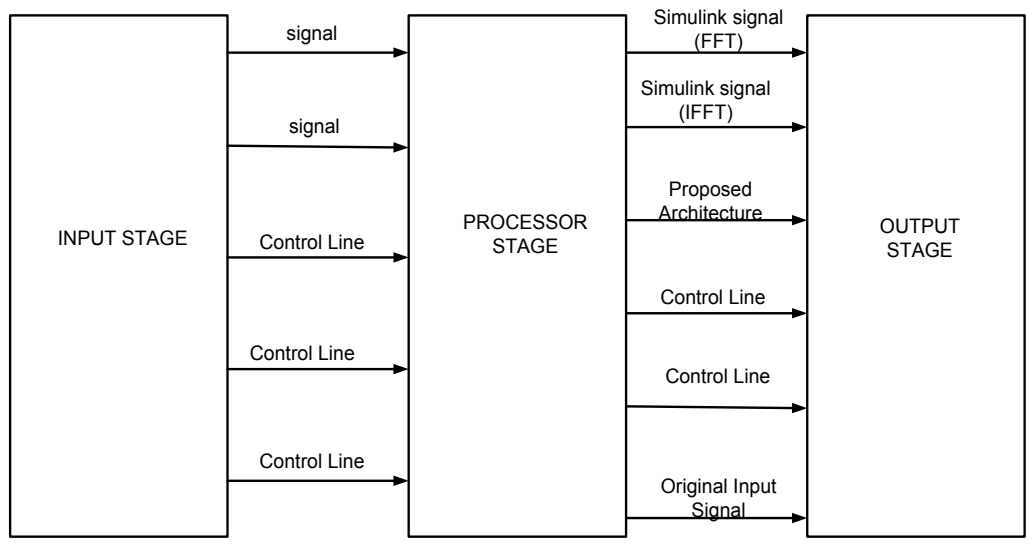

Fig. 2. Proposed pipelined system algorithm

## C. Pipeline Architecture

The same butterfly unit can perform the $N/2$ butterfly operations computed in every stage sequentially. Since the two inputs of a next butterfly unit of a stage are provided from the output of the butterfly unit of the previous stage at different time points, a shuffling unit is inserted between two successive butterfly units in order to route these outputs to the corresponding inputs of the next stage. To increase the system efficiency in the Radix butterfly algorithm, the pipeline registers are located after each addition, subtraction blocks that is the end of each stage. Hence, the pipeline butterfly algorithm keeps the final result in the register to be transferred to next step by the next calculation cycle. However the measurement of system efficiency after applying pipeline structure only can be evaluated after the hardware implementation. Fig. 3 shows the inner layer of proposed FFT processor design where pipelining is applied in signal input logic block.

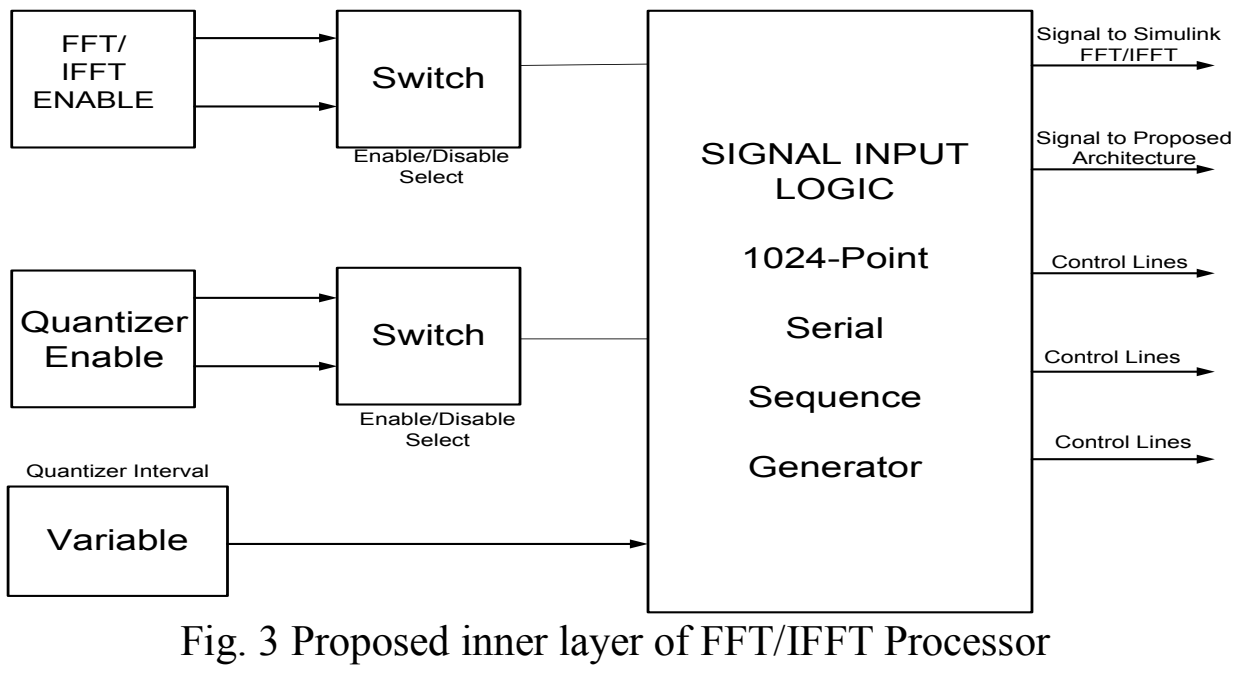

Fig. 3 Proposed inner layer of FFT/IFFT Processor

The signal input is inserted at the control signal to program the processor functionality. The control signals are to select FFT or IFFT calculation, while the other enables and disables the quantization of the twiddle factors. Fig. 4 illustrates the 10 stages butterfly for 1024-point pipeline FFT/IFFT processor.

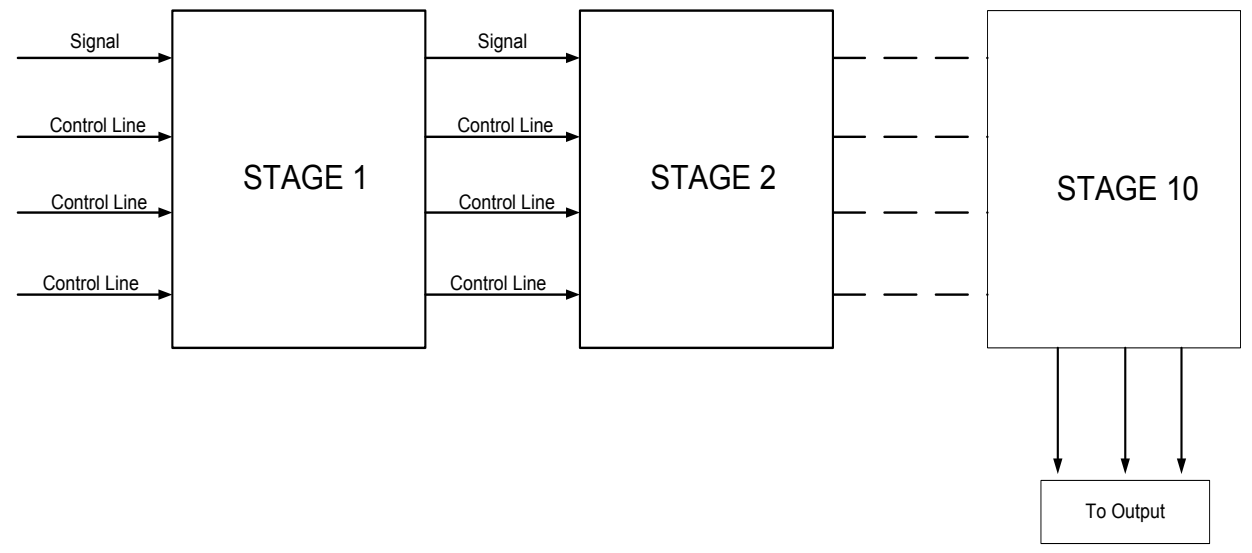

Fig. 4. Ten-Stage 1024-point pipeline processor

In the proposed design, excluding the input stage, the rest of stages consist of the twiddle factors, the shuffling unit and a floating-point quantize model. The interval of the quantize unit for each stage is preset statically and this is used to vary the bit-resolution of the processor. Fig. 5 shows the flowchart of overall system operation while quantizing and pipelining are applied.

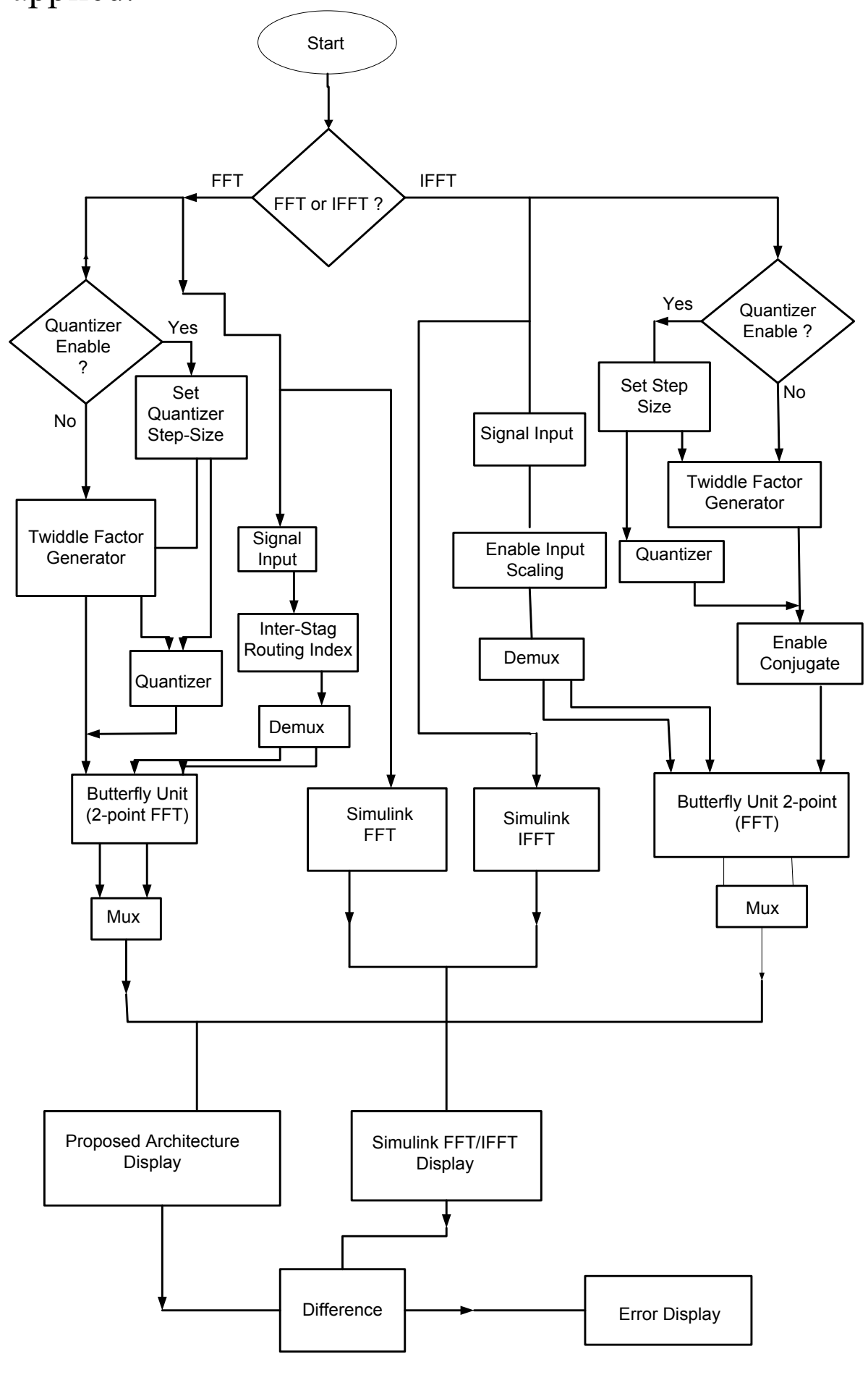

Fig. 5. Proposed overall system operation

The IFFT computation uses the same fundamental Radix-2 DIT Butterfly unit. However, the input is scaled by the factor of $N$ (1024). These discrete input values are then sent through the processor stage, which performs the same operation except that the conjugate of the twiddle factors are used instead. The output stage simply compares the results of the proposed FFT/IFFT Processor with the idle FFT/IFFT processor and their difference is observed as system error that will be analysed in the next chapter.

## III. Statistical Theory Of Quantization

### *A. Uniform Quantization*

One would expect that quantization has a similar effect on functions of the amplitude as sampling has on functions of time. Quantization is an operation on signals that is represented as a "staircase" function. Each input value is rounded toward the nearest allowable discrete level. The probability of each discrete output level equals the probability of the input signal occurring within the associated quantum band [16]. For example, the probability that the output signal has the value zero equals the probability that the input signal falls between $\pm q/2$, where $q$ is the quantization box size [8]. Fig. 6 shows the model of quantizing.

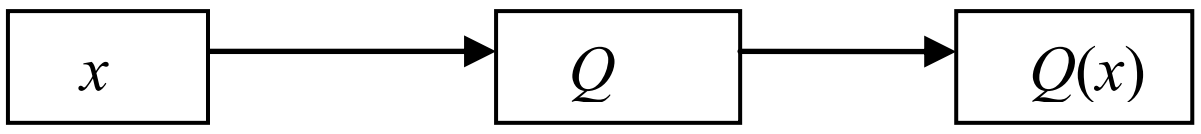

Fig. 6. Uniform Quantize Model

The quantize error ($h$) is given as

$$h = x - Q(x) \quad (3)$$

If $x$ and $h$ are real, with probability density function (PDF) as $P_x(.)$, then the quantization error variance is

$$\sigma_h^2 = E\{h^2\} = \int_{-\infty}^{\infty} h^2 P_h(h)dh$$
$$= \sum_{k=1}^{L} \int_{x_k}^{x_{k+1}} \left(x - Q(x)\right)^2 p_x(x)dx \quad (4)$$

$$\sigma_h^2 = \int_{-\frac{q}{2}}^{\frac{q}{2}} h^2 \frac{1}{q} dq = \frac{q^2}{12} = \frac{1}{3} x_{max}^2\, 2^{-2b} \quad (5)$$

$$SNR(dB) = 10log_{10}\left(\frac{\sigma_x^2}{\sigma_h^2}\right) \quad (6)$$

where $q$ is the quantization interval, $b$ is the number of bits. Quantization noise is defined as the difference between the output and input of the quantized signal. Since the quantized unit is designed in the proposed processor to enhance the calculations, Fig. 7 illustrates a plot of the error versus the number of bits for the uniform quantized, while Fig. 8 shows a comparison of the mean, standard deviation as well as variance of the uniform quantization model achieved by the proposed FFT Processor.

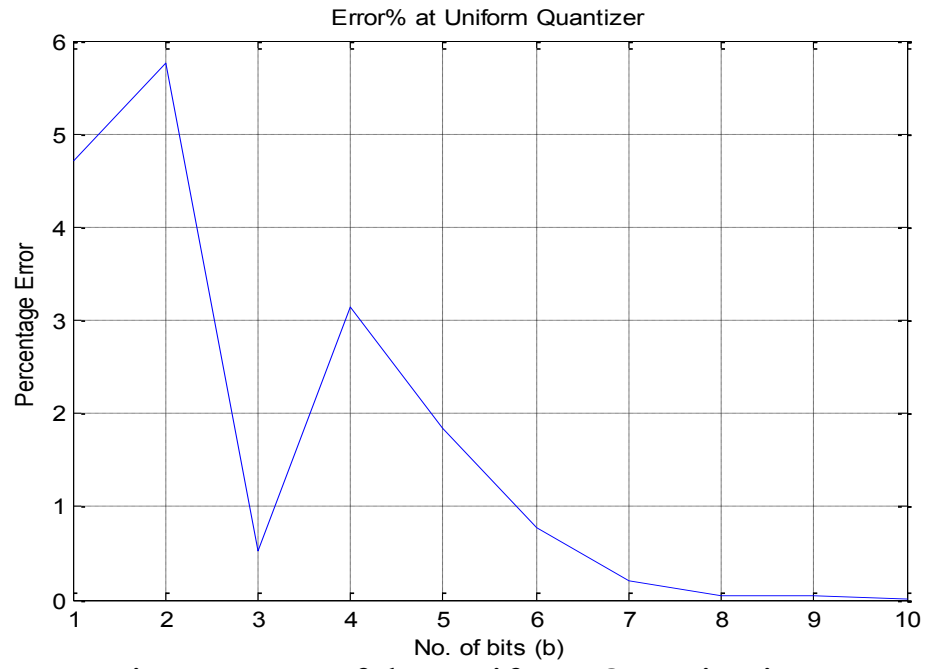

Fig. 7. Error of the Uniform Quantization

The probability of getting a given error value is the sum of probabilities of all the quantization boxes. The uniform quantize model performs uniform quantization on the signal input and thus, a linear signal is required at the input. Fixed-point numbers are considered linear since the Radix point remains fixed. However, this research was focused on floating-point numbers hence; the response of the uniform quantize to floating point input was observed. As result the quantization noise was increased. Equation (6) gives an expression for the SNR of the quantizer using the ratio of the variances of the input to noise.

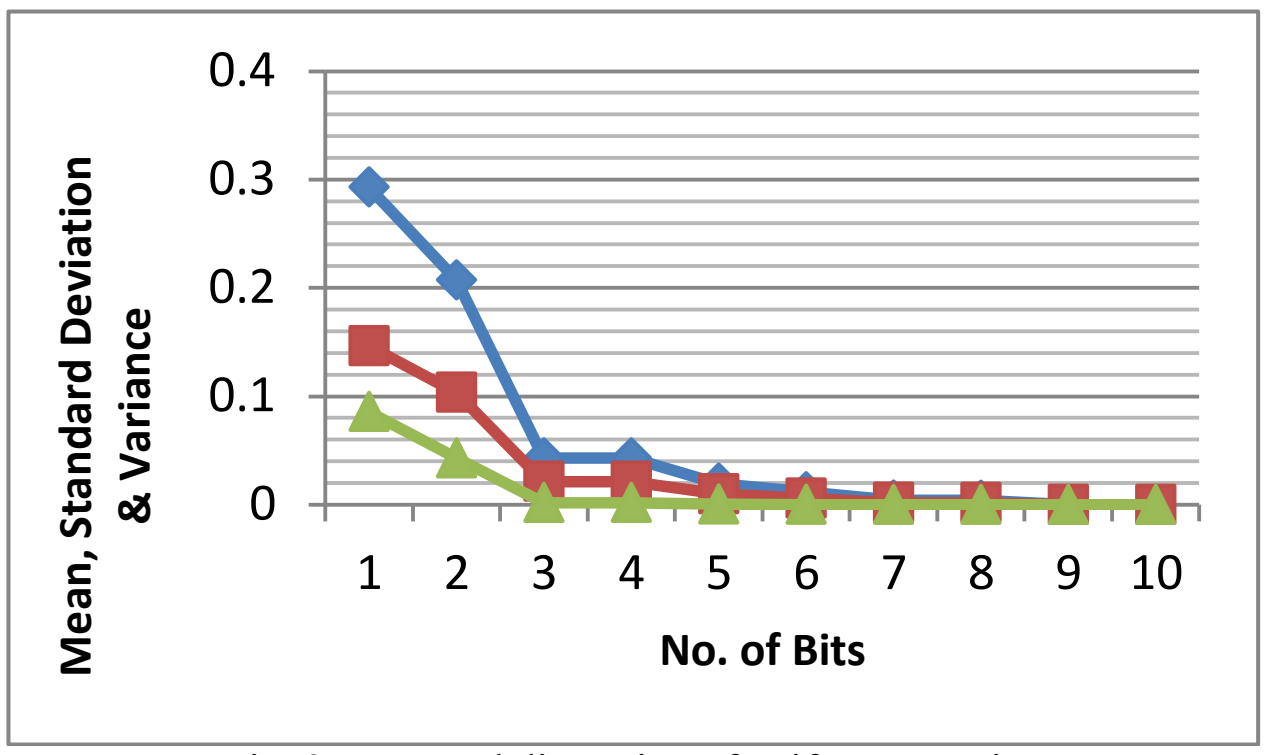

Fig. 8 Measured dispersion of uniform quantize

### *B. Non-Uniform Quantization*

The uniform quantization model is usually not used for floating-point quantization due to the overall non-uniform characteristic of the latter. Quantization of floating-point numbers is carried out only on the mantissa hence; it is more relevant to consider the relative error ε caused by the quantization process. The relative error defined in terms of the numerical values of the quantized floating-point number $Q(x) = 2^e Q(M)$ and the un-quantized number $x = 2^e M$ is given as

$$\varepsilon = \frac{(Q(x) - x)}{x} = \frac{(Q(M) - M)}{M} = \frac{\alpha}{M} \quad (7)$$

It is possible however, to represent the floating-point quantizer using a combination of a compressor, a uniform quantize and an expander. Fig. 9 shows the non-uniform quantized model.

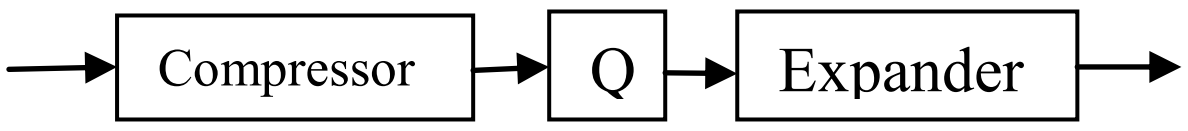

Fig. 9. Non-uniform quantization model

$$\sigma_\varepsilon^2 = E\{\varepsilon^2\} = \frac{2}{q} \int_{1/2}^{1} \int_{-q/2}^{q/2} \frac{\alpha^2}{M^2\, d\alpha} dM$$
$$= q^2/6 = (0.167)\, 2^{-2b} \quad (8)$$

An expression for the variance is shown in (8). The variance from the floating-point quantization equals half that obtained from the uniform quantization which is a generally preferred characteristic. Fig. 10 determines that the stability of the processor performance such that no variation occurs when the number of bits is 7 bit, unlike that of the uniform quantization which attains this stability at bit position eight 8, as shown in Fig. 7. That is the advantage of the system while modeling the floating-point structure.

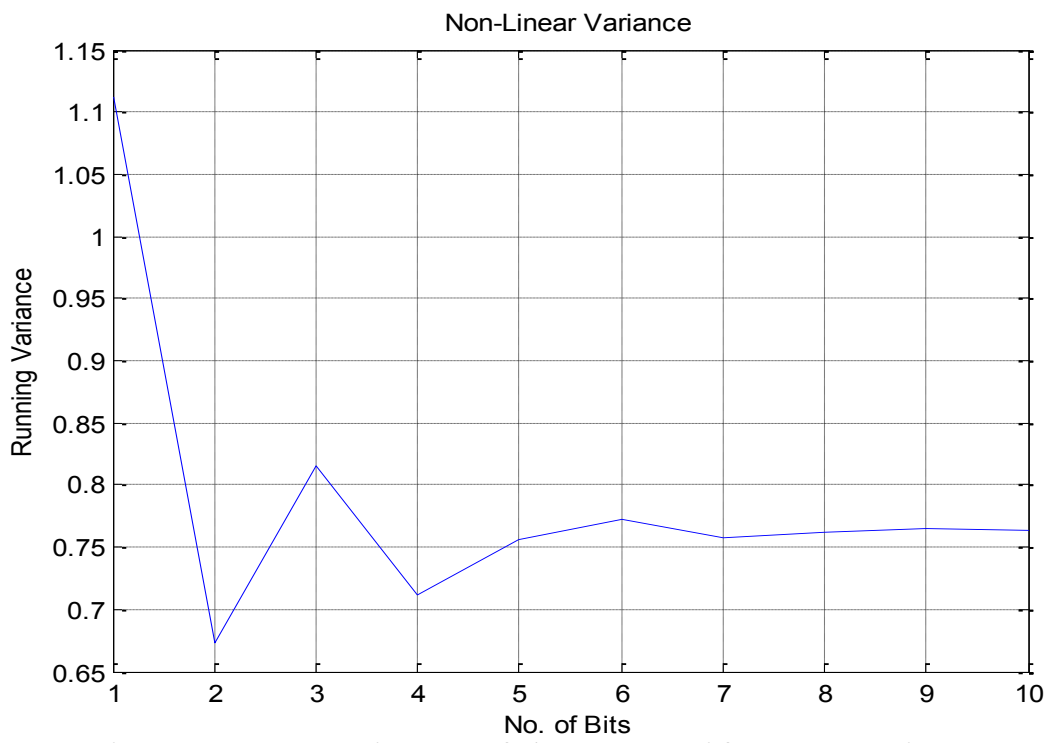

Fig.10 Error variance of the non-uniform quantizer

In addition, bit position 2 of Fig. 10 gives minimum swing before stability, contrary to that of Fig. 7 which occurs at bit position 3. This minimum swing gives a false minimum error position and can be used for less sensitive applications in which minimum error is not important.

Fig. 11 illustrates the comparison between mean, standard deviation when number of bit increased.

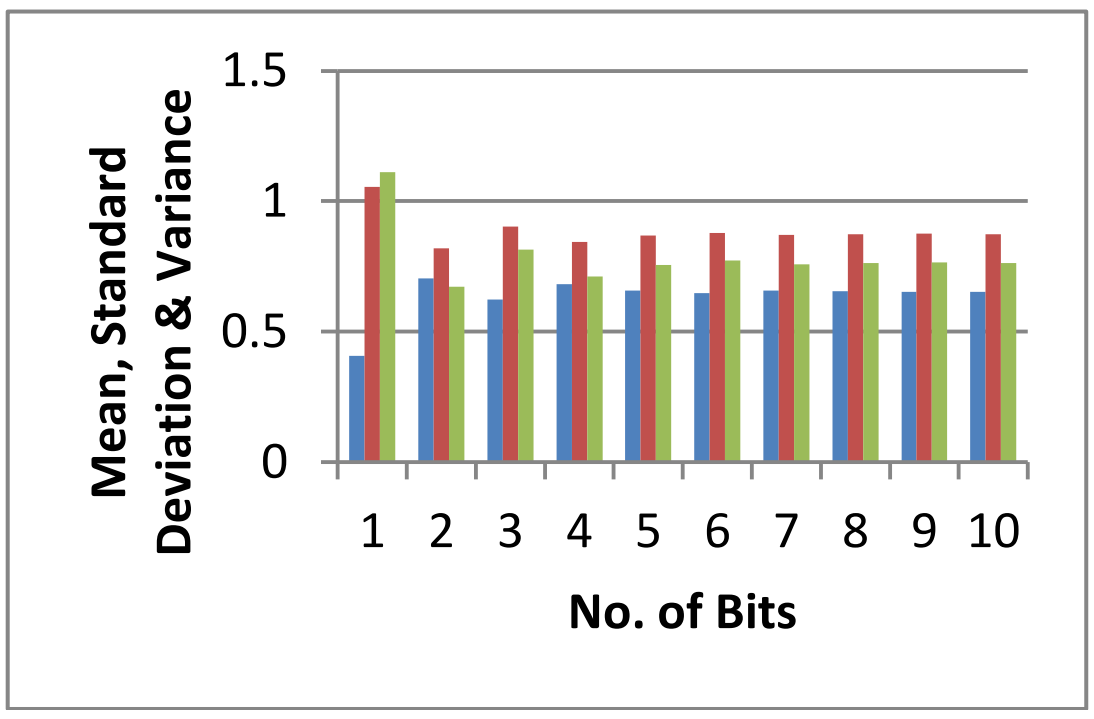

Fig. 11 Comparison of the measured dispersion of non-uniform quantization

## IV. Discussion

Statistical parameters like the mean, standard deviation and variance were used to analytically develop expressions for the variation in error as a function of the quantization interval. These parameters were also known to have relationships with the SQNR. The percentage error of the non-uniform quantization generally decreased with an increase in the quantizer interval.

As such, the SQNR is increased with respect to the quantization step size. The same general trend was observed in the uniform quantization, as well as the FFT and IFFT results. Fig. 12 shows the error variation when the input data increased.

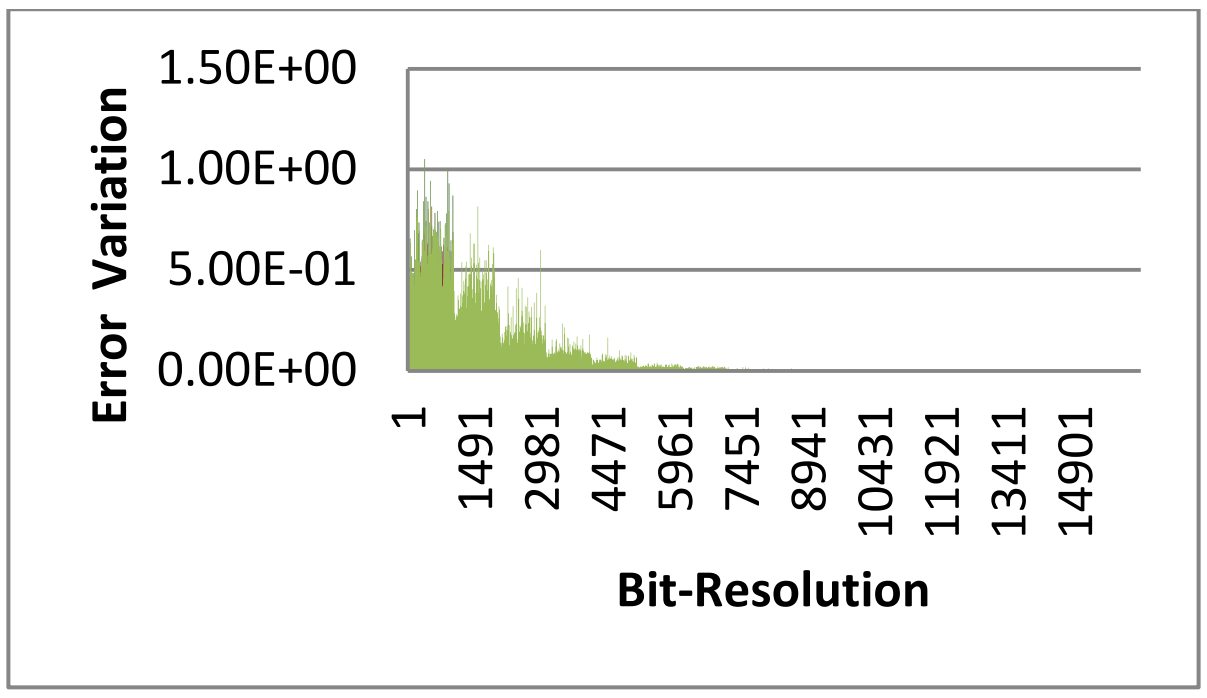

Fig. 12. Error variation of the FFT processor

The general trend observed from the results indicates that the measured dispersion can only be valuable when they are used alongside the mean since the mean actually provides the benchmark for understanding the decreasing trend. Fig. 13 shows the mean standard deviation and variance when the number of bit increased.

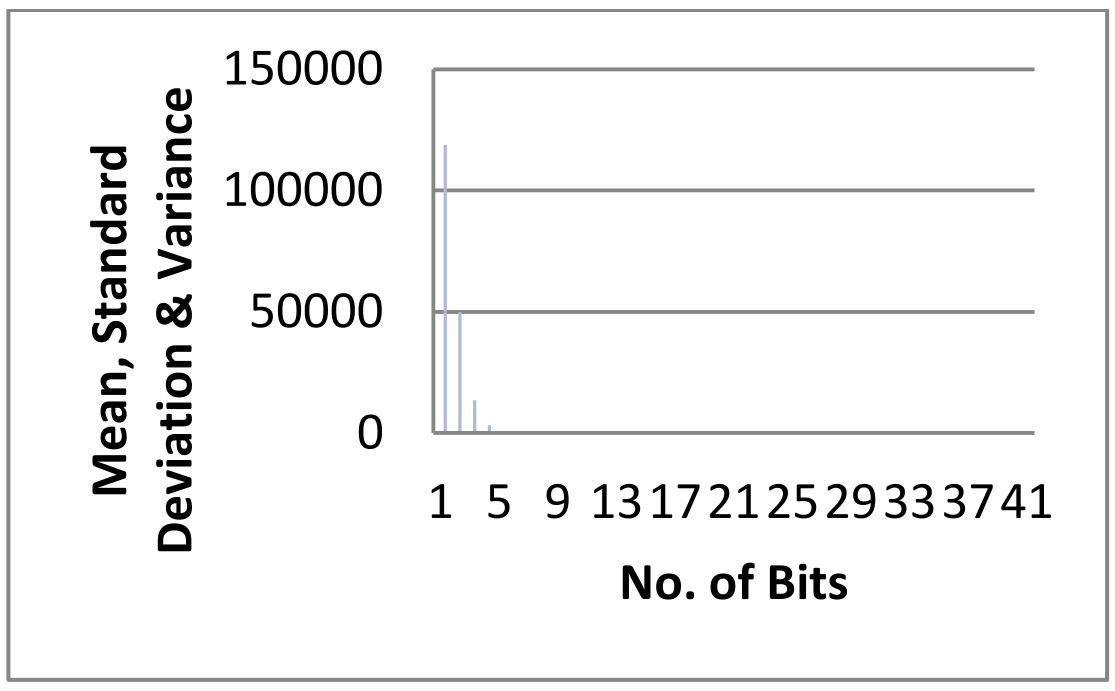

Fig. 13. Measured dispersion of FFT/IFFT processor

Therefore, the variance is decreased as quantization interval increased. Hence, the variance is inversely proportional to the percentage error, and as such, inversely proportional to the SQNR. This provides experimental proof to the theoretical models given earlier and provides a benchmark for the trade-off between the SQNR and the resolution. As shown in Fig. 14 the error variation also decreased as bit resolution increased.

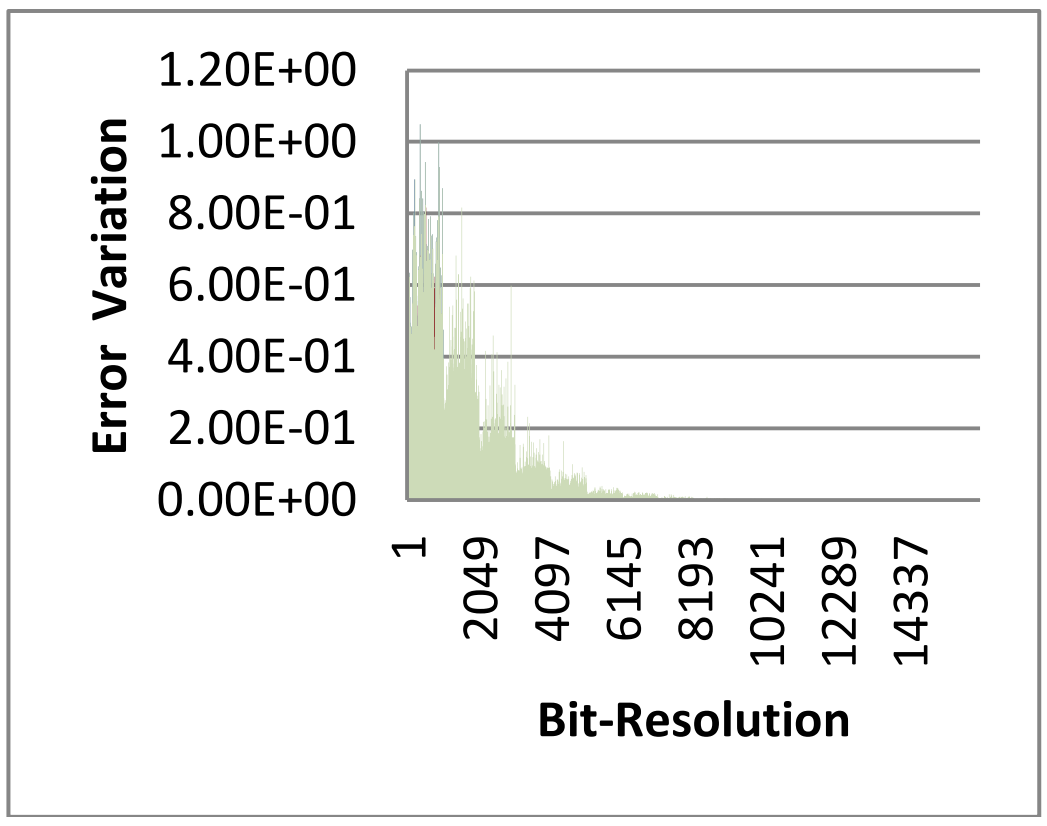

Fig. 14. Error variation of IFFT processor

## V. Conclusion

The 1024-point radix-2 FFT/IFFT processor is designed and simulated using MATLAB Simulink toolbox. The proposed processor satisfies the FFT size requirement of the 1024-points for a quantized pipeline structure. The percentage error and all the measured dispersion were found to decrease as the bit-resolution increased. This shows how the SQNR improves with bit-resolution. Although the power requirement for such SQNR systems are high, the proposed architecture provides an ease in the trade-off decision between the SQNR, power requirement and bit-resolution of the Radix-2 FFT/IFFT processor.